\documentclass[12pt]{iopart}
 \usepackage{iopams}
\newfont{\myeu}{eurm10 at 12 pt}
\newfont{\bfrak}{eufb10 at 12 pt}
\def\bN{\bar N}
\def\bK{\bar K}
\def\wb{{\overline W}}

\def\halft{\textstyle\frac 12}
\def\sfactor#1#2{\Bigg[\begin{array}{@{}c@{}}#1\\#2\end{array}\Bigg]}
\def\vq{^{\vphantom{+}}}
\begin{document}
\title[Eigenvectors in the Superintegrable Model II]%
{Eigenvectors in the Superintegrable Model II: Ground State Sector}

\author{Helen Au-Yang$^{1,2}$ and Jacques H H Perk$^{1,2}$%
\footnote{Supported in part by the National Science Foundation
under grant PHY-07-58139 and by the Australian Research Council
under Project ID: LX0989627.}}
\address{$^1$ Department of Physics, Oklahoma State University,
145 Physical Sciences, Stillwater, OK 74078-3072, USA%
\footnote{Permanent address.}}
\address{$^2$ Centre for Mathematics and
its Applications \& Department of Theoretical Physics,
Australian National University, Canberra, ACT 2600, Australia}
\ead{\mailto{perk@okstate.edu}, \mailto{helenperk@yahoo.com}}
\begin{abstract}
In 1993, Baxter gave $2^{m_Q}$ eigenvalues of the transfer
matrix of the $N$-state superintegrable chiral Potts model with
spin-translation quantum number $Q$, where
$m_Q=\lfloor(NL-L-Q)/N\rfloor$. In our previous paper we studied
the $Q=0$ ground state sector, when the size $L$ of the
transfer matrix is chosen to be a multiple of $N$. It was
shown that the corresponding $\tau_2$ matrix has a degenerate
eigenspace generated by the generators of $r=m_0$
simple ${\mathfrak{sl}}_2$ algebras. These results enable us to
express the transfer matrix in the subspace in terms of these
generators ${\mathbf E}_m^{\pm}$ and ${\mathbf H}\vq_m$
for $m=1,\cdots,r$. Moreover, the corresponding $2^r$ eigenvectors
of the transfer matrix are expressed in terms of rotated
eigenvectors of ${\bf H}\vq_m$.
\end{abstract}

\pacs{02.20.Uw, 05.50.+q, 64.60.De, 75.10.Hk, 75.10.Jm}
\vspace{2pc}


In our previous paper~\cite{APsu1}, we have shown that to calculate the
correlation functions of the chiral Potts model, one only needs to study
the eigenvectors of the superintegrable model. Since the transfer
matrix and the matrix $\tau_2(t_q)$ have the same eigenvectors, we have
examined the eigenspace of $\tau_2(t_q)$ which has degeneracy $2^r$,
where $r=L(N-1)/N$ with $L$---the length of the transfer
matrix---chosen to be a multiple of $N$. This space is a highest-weight
representation of an $\mathcal{L}({\mathfrak{sl}}_2)$ loop algebra
which turns out to be a direct sum of $r$ copies of the simple
algebra ${\mathfrak{sl}}_2$ whose generators are ${\bf E}_m^{\pm}$
and ${\bf H}\vq_m$ for $m=1,\cdots,r$.

In 1989, Baxter in his paper on the `Superintegrable Chiral Potts
Model'~\cite{Baxsu} suggested that the transfer matrix restricted to a
certain subspace may be written as
\begin{equation}
\mathcal{T}_Q=C_Q\rho^L\prod_{j=1}^r\{GI+(1-k'\cos\theta\vq_j)\sigma_j^z
-k'\sin\theta\vq_j\sigma_j^x\},
\end{equation}
(see Eq.\ (2.16) of ~\cite{Baxsu}). Our attempt is to do exactly that.
However, as the modulus $k'$ in that paper~\cite{Baxsu} is different
from the usual $k'$ of other papers, we shall consider the transfer
matrix of his later papers~\cite{BaxIf1,BaxIf2}.

{}From Eq.\ (6.2) of \cite{BaxIf1}\footnote{This simplifies under the
periodic boundary condition, i.e.\ $r=m_P=0$ in \cite{BaxIf1}
with $r$ being Baxter's skewness parameter here, not to be confused
with $r=L(N-1)/N$ in this paper.} the transfer matrices of the
superintegrable model can be written as
\begin{equation}
T_q=N^{{\frac 12}L}{\frac{(x_q-y_p)^L}{(x_q^N-y_p^N)^L}}
\mathcal{T}(x_q,y_q),\quad
{\hat T}_q=N^{{\frac 12}L}{\frac{(x_q-x_p)^L}{(x_q^N-x_p^N)^L}}
{\hat\mathcal{T}(x_q,y_q)}.
\label{ctcbt}\end{equation}
Baxter showed that the eigenvalues of $\mathcal{T}(x_q,y_q)$ for spin
shift quantum number $Q=0$, or equivalently $\mathcal{T}_0(x_q,y_q)$
corresponding to (I.10)\footnote{All equations in~\cite{APsu1} are
denoted here by prefacing I to its equation number.}, depend on
$\lambda_q=\mu_q^N$ only and are denoted by $\mathcal{G}(\lambda_q)$.
The corresponding eigenvalue $\hat\mathcal{G}(\lambda_q)$
of $\hat\mathcal{T}_0(x_q,y_q)$ can be made equal
to $\mathcal{G}(\lambda_q)$ by simple rescaling of the
eigenvectors $\mathbf{x}$ and $\mathbf{y}$, so that
\begin{equation}
\fl\mathcal{T}_0(x_q,y_q)\mathbf{x}=
\mathcal{G}(\lambda_q)\mathbf{y},\quad
\hat\mathcal{T}_0(x_q,y_q)\mathbf{y}=
\mathcal{G}(\lambda_q)\mathbf{x},\quad
\hat\mathcal{T}_0(y_q,x_q)\mathbf{y}=
\mathcal{G}(\lambda_q^{-1})\mathbf{x}.
\label{eveq}\end{equation}
Thus, (I.9) and (6.24) of \cite{BaxIf1} become\footnote{Eq.\
(I.11) with $Q=0$, $t\equiv t_q/t_p$ and (6.25) of \cite{BaxIf1}
with $P_a=P_b=0$, $F(t_q)\equiv1$ differ by a factor $N$.}
\begin{equation}
\mathcal{G}(\lambda\vq_q)\mathcal{G}(\lambda_q^{-1})
=t_p^{rN}N\mathcal{P}(t_q/t_p)
=t_p^{rN}N\prod_{j=1}^r\left[\left({{t_q}/{t_p}}\right)^N-z_j\right],
\label{gg}\end{equation}
where $z\vq_j$ and $z^{-1}_j$ are the roots of the Drinfeld polynomial
$P(z)=P(t^N)=\mathcal{P}(t)$, see (I.11) with $Q=0$ and (I.41).
From (I.2) we have
\begin{equation}
k^2 t_q^N=1+k'^2-k'(\lambda\vq_q+\lambda_q^{-1}).
\label{tld}\end{equation}
Substituting this equation into (\ref{gg}), and letting
\begin{equation}
2\cosh 2\theta\vq_j=k'+k'^{-1}-k^2 t_p^Nz\vq_j/k',
\label{theta}\end{equation}
we find
\begin{equation}
\mathcal{G}(\lambda\vq_q)\mathcal{G}(\lambda_q^{-1})
=N\Bigl(\frac{k'}{k^2}\Bigr)^r\prod_{j=1}^r\rme^{\mp2\theta_j}
(\rme^{\pm2\theta_j}-\lambda_q^{-1})
(\rme^{\pm2\theta_j}-\lambda\vq_q).
\label{gg2}\end{equation}
Therefore, as $\mathcal{G}(\lambda_q)$ is a polynomial
in $\lambda_q^{-1}$ according to section 6 of \cite{BaxIf1},
the $2^r$ implied eigenvalues of the transfer matrix
$\mathcal{T}_0(x_q,y_q)$ in (I.10) are
\begin{eqnarray}
&&\fl\mathcal{G}(\lambda_q)={\rho^r}\prod_{j=1}^r
[\cosh\theta\vq_j(1-\lambda_q^{-1})
\pm\sinh\theta\vq_j(1+\lambda_q^{-1})]
=\prod_{j=1}^r(A_j\pm B_j),
\label{eigent}\\
&&\fl A_j=\rho\cosh\theta\vq_j(1-\lambda_q^{-1}),\quad
B_j=\rho\sinh \theta\vq_j(1+\lambda_q^{-1}),\quad
\rho^r=N^{\frac12}({k'}/{k^2})^{{\frac12}r},
\label{abj}\end{eqnarray}
cf.\ (20) of \cite{BaxIf2}. These are the results of Baxter,
written in different forms.

We shall now attempt to express the transfer matrix
$\mathcal{T}_0(x_q,y_q)$ of (I.10) for $Q=0$ in terms of the
generators ${\bf E}_m^{\pm}$ and ${\bf H}\vq_m$, defined
for $m=1,\cdots,r$ in \cite{APsu1}, and also obtain the corresponding eigenvectors.

Similar to what Davies~\cite{Davies} did, we define the polynomials
\begin{equation}
f_j(z)=\prod_{\ell\ne j}{\frac{z-z_{\ell}}{z_j-z_{\ell}}}=
\sum_{n=0}^{r-1}\beta_{j,n}z^n, \quad
f_j(z_k)=\delta_{j,k}\,,
\label{beta}\end{equation}
where $\beta_{j,n}$ are the elements of the inverse of the Vandermonde
matrix~\cite{Prony}, such that
\begin{equation}
\sum_{n=0}^{r-1}\beta_{j,n}z_k^n=\delta_{j,k}, \qquad
\sum_{k=1}^{r}z_k^n\beta_{k,m}=\delta_{n,m}
\quad\hbox{for}\quad 0\le n\le r-1.
\label{vdm}\end{equation}
Consequently, (I.45)\footnote{We have replaced
${\bf E}_m^{\mp}\to\pm{\bf E}_m^{\pm}$ in (I.45), so that we obtain
the more usual (\ref{comEH}) rather than (I.46). Also, $z_m$
and $1/z_m$ are both zeroes of $P(z)$ when $Q=0$, because
of the symmetry of the coefficients of the polynomial $P(z)$ in (I.41),
so that (\ref{evdm}) and (I.45) are equivalent. The changes made seem
appropriate for the $Q\ne0$ case.} rewritten here as
\begin{equation}
\fl {\bf x}_n^{\mp}=\pm\sum_{m=1}^rz_m^{-n}{\bf E}_m^{\pm}=
\pm\sum_{m=1}^rz_{m^\ast}^n{\bf E}_m^{\pm}, \quad
{\bf h}\vq_n=\sum_{m=1}^rz_m^{-n}{\bf H}\vq_m=
\sum_{m=1}^rz_{m^\ast}^n{\bf H}\vq_m,
\label{evdm}\end{equation}
where $n\in\mathbb Z$ and $m^\ast$ is the subscript for
which $z_{m^\ast}=1/z_m$, can be inverted as
\begin{equation}
{\bf E}_m^{\pm}=
\pm\sum_{n=0}^{r-1}\beta\vq_{m^\ast,n}
z_m^{\,\ell}{\bf x}_{n+\ell}^{\mp}, \quad
{\bf H}\vq_m=
\sum_{n=0}^{r-1}\beta\vq_{m^\ast,n}z_m^{\,\ell}{\bf h}\vq_{n+\ell},
\label{eeh}\end{equation}
valid for $m=1,\ldots,r$ and $\ell\in\mathbb Z$.

In Appendix A, we shall show that
\begin{equation}
{\bf H}\vq_m|\Omega\rangle=-|\Omega\rangle,\qquad
({\bf E}_m^+)^2|\Omega\rangle=0,
\label{ho}\end{equation}
with $|\Omega\rangle$ the `ferromagnetic' ground state of \cite{APsu1},
whereas in Appendix B we shall give some further results for
${\bf E}_j^+{\bf E}_m^+|\Omega\rangle$.

Deguchi and Nishino have studied a similar loop algebra
\cite{Degu1, NiDe1,Degu2,NiDe2}. They have shown that there
exists a transformation to map the generators of the two loop
algebras \cite{NiDe2}. The proofs of the identities for the
related operators are also given in \cite{Degu2}.
(It can be verified that
our ${\bf E}_j^{\pm}$ is proportional to some $\rho_j^{\mp}({\bf a};s)$
of \cite{Degu2}, noting that Deguchi essentially uses Prony's method
\cite{Prony} to invert the Vandermonde matrix, but leaves out the
denominators in \cite{Prony}.) 

{}From the commutation relations, 
\begin{eqnarray}
\fl[{\bf E}_m^+,{\bf E}_n^-]=
\delta\vq_{m,n}{\bf H}\vq_m,\quad
[{\bf H}\vq_m,{\bf E}_n^{\pm}]=
\pm2\delta\vq_{m,n}{\bf E}_m^{\pm},\quad
[{\bf E}_m^+,{\bf E}_n^+]=[{\bf E}_m^-,{\bf E}_n^-]=0,
\label{comEH}\end{eqnarray}
cf.\ (I.46), we find
\begin{equation}
{\bf H}\vq_m{\bf E}_n^{+}|\Omega\rangle=
-(-1)^{\delta_{m,n}}{\bf E}_n^{+}|\Omega\rangle.
\label{heo}\end{equation}
Thus by operating with various products given by the subsets of
${\bf E}^+_1,{\bf E}^+_2,\cdots, {\bf E}^+_r$ on $|\Omega\rangle$, we
obtain the eigenvectors of ${\bf H}_m$, with eigenvalues $\xi_m$ and
$\xi_m=\pm 1$ depending on whether $m$ is in or not in the subset.
Following the example of Onsager~\cite{Onsager}, the $2^r$
eigenvectors are denoted by the eigenvalues $\xi_m$, namely,
\begin{equation}
\Psi(\xi_1,\xi_2,\cdots,\xi_r)=
\prod_{m\in J_n}{\bf E}_m^{+}|\Omega\rangle,
\quad \xi_j=\pm1,
\label{psi}\end{equation}
where $J_n=(j_1,\cdots,j_n)$ is any subset of $(1,2,\cdots,r)$ for
$n=0,\cdots, r$, with $\xi_j=1$ for
$j\in J_n$, and $\xi_j=-1$ for $j\notin J_n$. Equivalently,
\begin{equation}
{\bf H}_j\Psi(\xi_1,\xi_2,\cdots,\xi_r)=
\xi_j\Psi(\xi_1,\xi_2,\cdots,\xi_r).
\label{xp}\end{equation}
Particularly, for $n=0$ and $n=r$, we have
\begin{equation}
|\Omega\rangle=\Psi(-1,-1,\cdots,-1),\qquad
|\bar{\Omega}\rangle=\Psi(1,1,\cdots,1).
\label{xi}\end{equation}
Because $\xi_j=\pm 1$, it is easy to see that ${\bf H}_j^2={\bf 1}$ in
this restricted eigenspace. Since ${\bf E}_j^{-}|\Omega\rangle=0$, and
${\bf H}_j|\Omega\rangle=-|\Omega\rangle$, we find from (\ref{comEH})
\begin{eqnarray}
&&{\bf E}_j^{-}\Psi(\xi_1,\xi_2,\cdots,\xi_r)=
{\bf E}_j^{-}\prod_{m\in J_n}
{\bf E}_m^{+}|\Omega\rangle\nonumber\\
&&\hskip1in
=\left\{\begin{array}{ll}
\Psi(\xi_1,\cdots,\xi_{j-1},-\xi_j,\xi_{j+1},\cdots,\xi_r)&
\mbox{ if $j\in J_n$,}\\
0&\mbox{ if $j\notin J_n$.}\end{array}
\right.
\label{epo}\end{eqnarray}
Likewise, using $({\bf E}_j^{+})^2|\Omega\rangle=0$ we obtain
\begin{eqnarray}
&&{\bf E}_j^{+}\Psi(\xi_1,\xi_2,\cdots,\xi_r)=
{\bf E}_j^{+}\prod_{m\in J_n}
{\bf E}_m^{+}|\Omega\rangle\nonumber\\
&&\hskip1in=
\left\{\begin{array}{ll}
0&\mbox{ if $j\in J_n$,}\\
\Psi(\xi_1,\cdots,\xi_{j-1},-\xi_j,\xi_{j+1},\cdots,\xi_r)&
\mbox{ if $j\notin J_n$.}\end{array}
\right.
\label{empsi}\end{eqnarray}
Consequently, we find
\begin{equation}
({\bf E}_j^{+}+{\bf E}_j^{-})\Psi(\xi_1,\xi_2,\cdots,\xi_r)=
\Psi(\xi_1,\cdots,\xi_{j-1},-\xi_j,\xi_{j+1},\cdots,\xi_r),
\label{epmo}\end{equation}
and
\begin{equation}
({\bf E}_j^{+}+{\bf E}_j^{-})^2\Psi(\xi_1,\xi_2,\cdots,\xi_r)=
\Psi(\xi_1,\cdots,\xi_j,\cdots,\xi_r).
\label{e2pmo}\end{equation}
{}From the commutation relation (\ref{comEH}), we can show
\begin{equation}
[{\bf H}\vq_j,({\bf E}_j^{+}+{\bf E}_j^{-})]=
2({\bf E}_j^{+}-{\bf E}_j^{-}),\quad
[{\bf H}\vq_j,({\bf E}_j^{+}-{\bf E}_j^{-})]=
2({\bf E}_j^{+}+{\bf E}_j^{-}),
\label{comm1}\end{equation}
\begin{equation}
[({\bf E}_j^{+}-{\bf E}_j^{-}),({\bf E}_j^{+}+{\bf E}_j^{-})]=
2{\bf H}\vq_j,\quad
({\bf E}_j^{+}-{\bf E}_j^{-})^2=-1.
\label{comm2}\end{equation}

Now, similar to what Onsager did for the Ising model
in~\cite{Onsager}, we express the transfer matrix
$\mathcal{T}_0(x_q,y_q)$ in (I.10) within the eigenspace of
the eigenvalues given in (\ref{eigent}) as
\begin{eqnarray}
{\cal T}_0(x_q,y_q)={\bf \cal S}
\prod_{j=1}^r(A_j-{\bf H}_j B_j)
{\bf \cal R}^{-1},\label{transfer1} \\
\fl
{\hat{\cal T}}_0(x_q,y_q)={\bf \cal R}
\prod_{j=1}^r(A_j-{\bf H}_j B_j)
{\bf \cal S}^{-1},\quad
{\hat{\cal T}}_0(y_q,x_q)={\bf \cal R}
\prod_{j=1}^r({\bar A}_j-{\bf H}_j {\bar B}_j)
{\bf \cal S}^{-1},
\label{transfer23}\end{eqnarray}
where ${\bf \cal S}$ and ${\bf \cal R}$ are some complex rotation
operators, which shall be determined, and ${\bar A}_j$
and ${\bar B}_j$ in (\ref{transfer23}) can be obtained
from $A_j$ and $B_j$ in (\ref{abj}) by letting
$\lambda_q^{-1}\to\lambda\vq_q$.

If we introduce
\begin{equation}
|{\bf \cal X}_i\rangle={\bf \cal R}
\Psi(\xi_1,\xi_2,\cdots,\xi_r),\quad
|{\bf \cal Y}_i\rangle={\bf \cal S}
\Psi(\xi_1,\xi_2,\cdots,\xi_r),
\label{evector}\end{equation}
for $ i=1,\ldots,2^r$, then we find from (\ref{xp}) and
(\ref{transfer1}), cf.\ also (\ref{eveq}), that
\begin{equation}
\mathcal{T}_0(x_q,y_q)|{\bf \cal X}_i\rangle=
\mathcal{G}(\lambda_q,\{\xi_j\})|{\bf \cal Y}_i\rangle,\quad 
{\hat\mathcal{T}}_0(x_q,y_q)|{\bf \cal Y}_i\rangle=
\mathcal{G}(\lambda_q,\{\xi_j\})|{\bf \cal X}_i\rangle,
\label{eigeneq}\end{equation}
where the eigenvalues are
\begin{equation}
\mathcal{G}(\lambda_q,\{\xi_j\})=
\prod_{j=1}^r (A_j-\xi_j B_j).
\label{evalue}\end{equation}
Particularly, the ``largest" eigenvalue is
\begin{equation}
\mathcal{G}_0(\lambda_q)=\mathcal{G}(\lambda_q,\{-1\})=
\prod_{j=1}^r(A_j+B_j)
\label{eg0}\end{equation}
with ground-state eigenvector
\begin{equation}
|{\bf \cal X}_1\rangle={\bf \cal R}\Psi(-1,-1,...,-1)=
{\bf \cal R}|\Omega\rangle.
\label{evmax}\end{equation}

The rotations ${\bf \cal R}$ and ${\bf \cal S}$ should be chosen
such that they are independent of $q$ as the transfer matrices
with different $q$ commute. They are determined from the explicit
form of the transfer matrices given in (\ref{ctcbt}). We now calculate
the matrix elements of the transfer matrices. Using (I.4) and (I.10),
we rewrite the transfer matrices in (\ref{ctcbt}) (from configuration
$\{\sigma\}$ to $\{\sigma'\}$ in the row above) in terms of the edge
variables $\{n\vq_i\}$ and $\{n'_i\}$ with $n_i=\sigma_i-\sigma_{i+1}$
as
\begin{eqnarray}
\fl&&\langle\{n'_i\}|\mathcal{T}_Q(x_q,y_q)|\{n_i\}\rangle=
N^{-{\frac 12}L}\sum_{a=0}^{N-1}\omega^{-Qa}
\prod_{j=1}^L\Bigl[(y_p^N-x_q^N)}{(y\vq_p-x\vq_q)^{-1}\nonumber\\
\fl&&\qquad \times W_{pq}(a-N\vq_j+N'_j)
\wb_{p'q}(a-N\vq_{j+1}+N'_j)\Bigr],\qquad
N_j=\sum_{\ell<j} n_{\ell}=\sigma_1-\sigma_j,
\label{otrans1}\end{eqnarray}
where $a=\sigma_1-\sigma'_1$, and 
\begin{equation}
\fl W_{pq}(n)={\Bigl(\frac{\mu_p}{\mu_q}\Bigr)}^{n}
\prod^{n}_{j=1}\frac{y_q-x_p\omega^j}{ y_p-x_q\omega^j},\quad
\wb_{p'q}(n)={\Bigl(\frac{\mu_q}{\mu_p}\Bigr)}^n
\prod^{n}_{j=1}\frac{\omega y_p-x_q\omega^j}{ y_q-x_p\omega^j}.
\label{wwb}\end{equation}
Cancelling out the common factors in the weights, (\ref{otrans1})
becomes
\begin{eqnarray}
&&\fl\langle\{n'_i\}|\mathcal{T}_Q(x_q,y_q)|\{n_i\}\rangle=
N^{-{\frac 12}L}\sum_{a=0}^{N-1}\omega^{-Qa}\nonumber\\
&&\fl\qquad\times\prod_{j=1}^L\Biggl[{\frac{(y_p^N-x_q^N) 
\omega^{a-N\vq_{j+1}+N'_j}}
{y_p-x_q\omega^{a-N\vq_{j+1}+N'_j}}}
{\Bigl(\frac{\mu_p}{\mu_q}\Bigr)}^{n_j}\prod^{n_j}_{\ell=1}
\frac{y_q-x_p\omega^{\ell+a-N\vq_{j+1}+N'_j}}
{y_p-x_q\omega^{\ell+a-N\vq_{j+1}+N'_j}}\Biggr].
\label{nntrans1}\end{eqnarray}
It is easy to see that for the two ground states,
\begin{equation}
\fl|\Omega\rangle\;\leftrightarrow\;n_i\equiv N_i
\equiv0\quad\hbox{or}
\quad|\bar\Omega\rangle\;\leftrightarrow\;n_i\equiv N-1,\;
N_i\equiv(i-1)(N-1),
\end{equation}
the above expression simplifies to
\begin{equation}
\langle\{n_i\}|\mathcal{T}_Q(x_q,y_q)|\Omega\rangle
=N^{-{\frac 12}L}\sum_{a=0}^{N-1}\omega^{-Qa}
\prod_{j=1}^L
\frac{\omega^{a+N_j}(y_p^N-x_q^N)}
{y_p-x_q\omega^{a+N_j}},
\label{nto}\end{equation}
and
\begin{equation}
\fl
\langle\{n_i\}|\mathcal{T}_Q(x_q,y_q)|{\bar\Omega}\rangle
=N^{-{\frac 12}L}\sum_{a=0}^{N-1}\omega^{-Qa}
\prod_{j=1}^L {\Bigl(\frac{\mu_p}{\mu_q}\Bigr)}^{N-1}
\frac{\omega^{a+j+N_j}(y_q^N-x_p^N)}
{y_q-x_p\omega^{a+j+N_j}},
\label{ntbo}\end{equation}
where also $n_1+\cdots+n_L\equiv0\, (\hbox{mod}\,N)$, as is required
by the periodic boundary conditions. For (\ref{ntbo}) we have used
$\prod_{\ell=1}^N(y-x\omega^{\ell})=y^N-x^N $. Particularly, when
$n_1=\cdots=n_L=0$, we find $N_j=0$ and (\ref{nto}) becomes
\begin{equation}
\langle \Omega|\mathcal{T}_Q(x_q,y_q)|\Omega\rangle
=N^{1-{\frac 12}L}y_p^{rN}(x_q/y_p)^Q\mathcal{P}_Q(x_q/y_p),
\label{oto2}\end{equation}
while for $n_1=\cdots=n_L=N-1$, we substitute $N_j=(j-1)(N-1)$
into (\ref{nto}) to obtain
\begin{equation}
\langle{\bar\Omega}|\mathcal{T}_Q(x_q,y_q)|\Omega\rangle
=N^{1-{\frac 12}L}\delta_{Q,0}\omega^{-L(L+1)/2}(y_p^{N}-x_q^N)^r.
\label{boto}\end{equation}
Similarly we find from (\ref{ntbo}) 
\begin{equation}
\langle{\bar\Omega}|\mathcal{T}_Q(x_q,y_q)|{\bar\Omega}\rangle
=N^{1-{\frac 12}L}\omega^Q(\mu_p y_q/\mu_q)^{rN}(x_p/y_q)^Q
\mathcal{P}_Q(x_p/y_q)
\label{botbo2}\end{equation}
and
\begin{equation}
\langle \Omega|\mathcal{T}_Q(x_q,y_q)|{\bar\Omega}\rangle
=N^{1-{\frac 12}L}\delta_{Q,0}\omega^{L(L+1)/2}(y_p^{N}-x_q^N)^r.
\label{otbo}\end{equation}
Since $L$ is a multiple of $N$, the right-hand-sides of (\ref{boto})
and (\ref{otbo}) are equal, this means that the signs of ${\bf E}^+_j$
and ${\bf E}^-_j$ are equal, and that the transfer matrix can only
have the following form for the $2^r$-dimensional ground state sector:
\begin{eqnarray}
{\cal T}_0(x_q,y_q)=\prod_{j=1}^r[X\vq_j-{\bf H}\vq_j Y\vq_j+
({\bf E}^+_j +{\bf E}^-_j)Z\vq_j].
\label{transfer1a}\end{eqnarray}

To evaluate $X_j$, $Y_j$ and $Z_j$, we need to evaluate
$\langle\Omega|{\bf E}_m^{-}$ and $\langle{\bar\Omega}|{\bf E}_m^{+}$.
To do so, we invert both (I.42) and
\begin{eqnarray}
\fl\langle\Omega|({\bf x}_0^+)^{(n)}({\bf x}_{1}^-)^{(n-1)}=
\sum_{j=1}^n\Lambda_{n-j}\langle\Omega|{\bf x}_{j-1}^+,\quad
\langle{\bar \Omega}|({\bf x}_1^-)^{(n)}({\bf x}_{0}^+)^{(n-1)}=
\sum_{j=1}^n\Lambda_{n-j}\langle{\bar\Omega}|{\bf x}_{j}^-,
\end{eqnarray}
which can be derived similarly, as
\begin{eqnarray}
\fl{\bf x}_j^-|\Omega\rangle=
\sum_{n=1}^j S_{j-n}({\bf x}_{0}^+)^{(n-1)}
({\bf x}_1^-)^{(n)}|\Omega\rangle,\quad
{\bf x}_{j-1}^+|{\bar\Omega}\rangle=
\sum_{n=1}^j S_{j-n}({\bf x}_{1}^-)^{(n-1)}
({\bf x}_0^+)^{(n)}|{\bar\Omega}\rangle,\\
\fl\langle\Omega|{\bf x}_{j-1}^+=
\sum_{n=1}^j S_{j-n}\langle\Omega|
({\bf x}_0^+)^{(n)}({\bf x}_{1}^-)^{(n-1)},\quad
\langle{\bar\Omega}|{\bf x}_{j}^-=
\sum_{n=1}^j S_{j-n}\langle{\bar \Omega}|
({\bf x}_1^-)^{(n)}({\bf x}_{0}^+)^{(n-1)},
\label{inxpm}\end{eqnarray}
where
\begin{equation}
\sum_{n=0}^{m}\Lambda_{m-n}S_n=\delta_{m,0},\quad \hbox{with}
\quad S_0=1,\quad\Lambda_{0}=1.
\label{S}\end{equation}
This is most easily seen by viewing $\mathbf\Lambda$ as a lower
triangular $r+1$ by $r+1$ matrix with elements
$\mathbf\Lambda_{i,j}=\Lambda_{i-j}$, zero whenever $j>i$, and
a similar form for its inverse. Letting
\begin{equation}
P(z)=\sum_{n=0}^{r}\Lambda_{n}z^n, \quad
Q(z)=\sum_{n=0}^{r}S_{n}z^n,
\label{cpq}\end{equation}
we find from (\ref{S}) that
\begin{equation}
Q(z)=1/P(z)=\prod_{n=1}^{r}(z-z_n)^{-1}=
\sum_{i=1}^{r}(z-z_i)^{-1}\prod_{\ell\ne i}^{r}(z_i-z_\ell)^{-1}.
\label{qip}\end{equation}
Using (\ref{beta}) with $z=0$ together with
$\prod_{\ell=1}^r(-z_\ell)=1$, we obtain
\begin{equation}
S_n=\sum_{i=1}^{r}z_i^{-n}\beta_{i,0},\quad
\hbox{for }1\!-\!r<n<r\!-\!1,
\label{sb}\end{equation}
also using (\ref{vdm}), which implies $S_n=0$ for $1\!-\!r\!<n\!<0$.
Substituting (\ref{inxpm}) into (\ref{eeh}) with $\ell=0$,
interchanging the order of summation, and using (\ref{sb}), we get
\begin{equation}
\langle \Omega|{\bf E}_m^{-}=-\langle \Omega|\sum_{\ell=1}^r
({\bf x}_0^+)^{(\ell)}({\bf x}_1^-)^{(\ell-1)}
\sum_{i=1}^r\beta_{i,0}\sum_{j=0}^{r-1}
\beta_{m^\ast,j}z_i^{\ell-j-1}.
\label{oep}\end{equation}
where the condition $ S_n=0$ for negative values of $n$ has been
used to extend the interval of the $j$ summation. As (I.41) implies
the symmetry $P(z)=z^rP(1/z)$, we find that for each root $z_i$ of
$P(z)$, $z_{i^*}\equiv1/z_i$ is also a root. We now use the first
equation in (\ref{vdm}), i.e.
\begin{equation}
\sum_{j=0}^{r-1}\beta_{m^\ast,j}z_i^{-j}=
\sum_{j=0}^{r-1}\beta_{m^\ast,j}z_{i^\ast}^j=\delta_{m,i}\,,
\label{oepd}\end{equation}
to get
\begin{eqnarray}
\langle\Omega|{\bf E}_m^{-}=
-\beta_{m,0}\sum_{\ell=1}^r z_m^{\ell-1}
\langle\Omega|({\bf x}_0^+)^{(\ell)}({\bf x}_1^-)^{(\ell-1)},
\nonumber\\
{\bf E}_m^-|{\bar\Omega}\rangle=
-\beta_{m,0}\sum_{\ell=1}^r z_m^{\ell-1}
({\bf x}_1^-)^{(\ell-1)}({\bf x}_0^+)^{(\ell)}|{\bar\Omega}\rangle.
\label{oep2}\end{eqnarray}
Similarly, but now using  (\ref{eeh}) with $\ell=1$, we can show
\begin{eqnarray}
{\bf E}_m^+|\Omega\rangle=
\beta_{m,0}\sum_{\ell=1}^r z_m^{\ell}
({\bf x}_0^+)^{(\ell-1)}({\bf x}_1^-)^{(\ell)}|\Omega\rangle,
\nonumber\\
\langle{\bar \Omega}|{\bf E}_m^{+}=
\beta_{m,0}\sum_{\ell=1}^r z_m^{\ell}
\langle{\bar\Omega}|({\bf x}_1^-)^{(\ell)}({\bf x}_0^+)^{(\ell-1)}.
\label{emo}\end{eqnarray}
{}From (I.20), we find\footnote{We use the standard definition
$\left[{{n'}\atop{n}}\right]\equiv\frac{[n']!}{[n]![n'-n]!}$.}
\begin{eqnarray}
\frac{{\mbox{\bfrak e}}^n}{[n]!} |n'\rangle=
\sfactor{n'}{n}|n'-n\rangle,\quad
&&\frac{{\mbox{\bfrak f}}^n}{[n]!} |n'\rangle=
\sfactor{n'+n}{n}|n'+n\rangle,\nonumber\\
\langle n'|\frac{{\mbox{\bfrak e}}^n}{[n]!} =
\sfactor{n'+n}{n}\langle n'+n|,\quad
&&\langle n'|\frac{{\mbox{\bfrak f}}^n}{[n]!}=
\sfactor{n'}{n}\langle n'-n|.
\label{oefo}\end{eqnarray}
Using (I.35)\footnote{There are some misprints in (I.35): The four
subscripts $j$ must be replaced by $m$.} and the above formulas,
we obtain
\begin{eqnarray}
({\bf x}_1^-)^{(\ell)} |\Omega\rangle=
\sum_{ {\{0\le n_j\le N-1\}}\atop{n_1+\cdots+n_L=\ell N}}
\omega^{-\sum_{j}j n_j}|\{n_j\}\rangle,\nonumber\\
({\bf x}_0^+)^{(\ell)} |{\bar\Omega}\rangle=(-1)^\ell
\sum_{ {\{0\le n_j\le N-1\}}\atop{n_1+\cdots+n_L=\ell N}}
|\{N-1-n_j\}\rangle,
\label{xmo1}\\
\langle \Omega|({\bf x}_0^+)^{(\ell)}=
\sum_{ {\{0\le n_j\le N-1\}}\atop{n_1+\cdots+n_L=\ell N}}
\omega^{\sum_{j} j n_j}\langle \{n_j\}|,\nonumber\\
\langle{\bar \Omega}|({\bf x}_1^-)^{(\ell)}=(-1)^\ell
\sum_{ {\{0\le n_j \le N-1\}}\atop{n_1+\cdots+n_L=\ell N}}
\langle\{N-1-n_j\}|,
\label{xmo}\end{eqnarray}
where we have twice used
\begin{equation}
\sfactor{N-1-n}{\nu}=(-1)^{\nu}\omega^{-n\nu-\nu(\nu+1)/2}
\sfactor{\nu+n}{\nu}
\label{sfactor}\end{equation}
with $n=0$. Continuing the same process we find
\begin{eqnarray}
({\bf x}_0^+)^{(\ell)}({\bf x}_1^-)^{(\ell+1)} |\Omega\rangle
=\sum_{ {\{0\le n_j\le N-1\}}\atop{n_1+\cdots+n_L=N}}
\omega^{-\sum_{j}j n_j}K_{\ell N}(\{n_j\})
|\{n_j\}\rangle,\nonumber\\
\langle \Omega|({\bf x}_0^+)^{(\ell+1)}({\bf x}_1^-)^{(\ell)}
=\sum_{ {\{0\le n_j\le N-1\}}\atop{n_1+\cdots+n_L=N}}
\langle\{n_j\}|\omega^{\sum_{j} j n_j}{\bar K}_{\ell N}(\{n_j\}),
\nonumber\\
({\bf x}_1^-)^{(\ell)}({\bf x}_0^+)^{(\ell+1)} |\bar\Omega\rangle
=-\sum_{ {\{0\le n_j\le N-1\}}\atop{n_1+\cdots+n_L=N}}
K_{\ell N}(\{n_j\})
|\{N-1-n_j\}\rangle,\nonumber\\
\langle{\bar \Omega}|({\bf x}_1^-)^{(\ell+1)}({\bf x}_0^+)^{(\ell)}
=-\sum_{ {\{0\le n_j\le N-1\}}\atop{n_1+\cdots+n_L=N}}
\langle\{N-1-n_j\}|{\bar K}_{\ell N}(\{n_j\}),
\label{xpmo}\end{eqnarray}
where 
\begin{eqnarray}
&&K_m(\{n_j\})\equiv
\sum_{ {\{0\le n'_j\le N-1\}}\atop{n'_1+\cdots+n'_L=m} }
\prod_{j=1}^L\sfactor{n\vq_j+n'_j}{n'_j}\omega^{n'_j N\vq_j},
\quad N_j\equiv\sum_{\ell=1}^{j-1}n_\ell\,, \label{sumK}\\
&&\bK_m(\{n_j\})\equiv
\sum_{{\{0\le n'_j\le N-1\}}\atop{n'_1+\cdots+n'_L=m}}
\prod_{j=1}^L\sfactor{n\vq_j+n'_j}{n'_j}\omega^{n'_j\bN\vq_j},
\quad \bN_j\equiv\sum_{\ell=j+1}^{L}n_\ell\,,
\label{sumbK}\end{eqnarray}
for integers $m\le(N-1)L$. Also we have
$\sum_jn'_j N\vq_j=\sum_j n\vq_j\bN'_j$,
$\sum_jn'_j\bN\vq_j=\sum_j n\vq_jN'_j$.
The generating function of $K_{m}(\{n_j\})$ for $n_1+\cdots+n_L=kN$ is 
\begin{equation}
g(\{n_j\},t)\equiv\sum_{m=0}^{(N-1)L-k N}K_m(\{n_j\})t^m=
\frac 1{(1-t^N)^{k}}\prod_{j=1}^{L}\frac{1-t^N}{1-t\omega^{N_j}},
\label{ck1}\end{equation}
which is obtained by inserting $t^{n'_j}$ in each of the $L$ sums of
$K_{m}(\{n_j\})$ and summing over $m$ to remove the constraint;
after using (\ref{sfactor}) and other $\omega$-binomial identities we
arrive at (\ref{ck1}). Similarly, we can define $\bar g(\{n_j\},t)$
with $K_m(\{n_j\})$ replaced by $\bar K_m(\{n_j\})$ and $N_j$
by $\bar N_j$. As seen from (\ref{sumK}) and (\ref{sumbK}), here we
only need to consider the case with $k=1$.

We define the following polynomials for $n_1+\cdots+n_L=N$,
\begin{eqnarray}
\fl&&G_Q(\{n_j\},z)\equiv\sum_{\ell=0}^{m_Q-1}
K_{Q+\ell N}(\{n_i\})z^{\ell}=\frac{t^{-Q}
}{N(1-t^N)}\sum_{a=0}^{N-1}\omega^{-Qa}
\prod_{j=1}^L\frac {1-t^N}{1-t\omega^{a+N_j}},
\label{ck2}\end{eqnarray}
and
\begin{eqnarray}
\fl&&
\bar{G}_Q(\{n_j\},z)\equiv\sum_{\ell=0}^{m_Q-1}
{\bar K}_{Q+\ell N}(\{n_j\})z^{\ell}=\frac{t^{-Q}}
{N(1-t^N)}\sum_{a=0}^{N-1}\omega^{-Qa}
\prod_{j=1}^L\frac{1-t^N}{1-t\omega^{a+\bN_j}},
\label{chk2}\end{eqnarray}
where $z=t^N$. The last equalities in (\ref{ck2}) and (\ref{chk2})
are proved noting
\begin{equation}
G_Q(\{n_j\},z)=N^{-1}\sum_{a=0}^{N-1}(t\omega^a)^{-Q}
g(\{n_j\},t\omega^a),
\end{equation}
and a similar equation for $\bar G_Q(\{n_j\},z)$.
For $Q=0$, we drop the subscripts in $G_0(\{n_j\},z)$ and
$\bar G_0(\{n_j\},z)$.
It is easy to see that $\bar{G}(\{n_j\},z)$ is the complex
conjugate of $G(\{n_j\},z)$ when $t$ is real.

Substituting (\ref{xpmo}) into (\ref{oep2}) and (\ref{emo}), and using
(\ref{ck2}) and (\ref{chk2}), we obtain
\begin{eqnarray}
\langle \Omega|{\bf E}_m^{-}=-\beta_{m,0}
\sum_{ {\{0\le n_j\le N-1\}}\atop{n_1+\cdots+n_L=N}} \langle \{n_j\}|\,
\omega^{\sum_{j} j n_j}\bar{G}(\{n_j\},z_m),
\label{oep3}\\
\langle{\bar \Omega}|{\bf E}_m^{+}=-\beta_{m,0}z_m
\sum_{ {\{0\le n_j\le N-1\}}\atop{n_1+\cdots+n_L=N}}
\langle \{N-1-n_j\}|\,
\bar{G}(\{n_j\},z_m)
,\label{boem}\\
{\bf E}_k^{+}|\Omega\rangle=\beta_{k,0}z_k
\sum_{ {\{0\le n_j\le N-1\}}\atop{n_1+\cdots+n_L=N}}
\omega^{-\sum_{j} j n_j}
{G}(\{n_j\},z_k)|\{n_j\}\rangle,\label{emo2}\\
{\bf E}_k^{-}|{\bar\Omega}\rangle=\beta_{k,0}
\sum_{ {\{0\le n_j\le N-1\}}\atop{n_1+\cdots+n_L=N}}
{G}(\{n_j\},z_k)|\{N-1-n_j\}\rangle.
\label{epbo}\end{eqnarray}
Since
\begin{equation}
\langle \Omega|{\bf E}_m^{-}{\bf E}_k^{+}|\Omega\rangle=\delta_{m,k},
\end{equation}
we conclude that
\begin{equation}
\beta_{m,0} \beta_{k,0}z_{k}
\sum_{ {\{0\le n_j\le N-1\}}\atop{n_1+\cdots+n_L=N}}
\bar{G}(\{n_j\},z_{m})G(\{n_j\},z_{k})
=-\delta_{mk}\,.
\label{ortho}\end{equation}
Next, we introduce the polynomials
\begin{equation}
{\mbox{\myeu h}}_k(z)\equiv
\sum_{ {\{0\le n_j\le N-1\}}\atop{n_1+\cdots+n_L=N}}
\bar{G}(\{n_j\},z_k)G(\{n_j\},z).
\label{dh}\end{equation}
Since the degree of polynomial $G(\{n_j\},z)$ defined in (\ref{ck2})
is $r-1$, the degree of ${\mbox{\myeu h}}_k(z)$ is also $r-1$. From
(\ref{ortho}) we know its $r-1$ roots; thus ${\mbox{\myeu h}}_k(z)$
is proportional to $f_k(z)$ defined in  (\ref{beta}). In fact,
we have
\begin{equation}
{\mbox{\myeu h}}_k(z)=-f\vq_k(z)/z\vq_k\beta_{k,0}^2=
\beta_{k,0}^{-1}\prod_{\ell\ne k}(z-z_\ell),
\label{hf}\end{equation}
where we have used $f_k(0)=\beta_{k,0}$
and $\prod_{\ell=1}^r(-z_\ell)=1$. Similarly, we have
\begin{equation}
{\bar{\mbox{\myeu h}}}_k(z)\equiv
\sum_{ {\{0\le n_j\le N-1\}}\atop{n_1+\cdots+n_L=N}}
\bar{G}(\{n_j\},z)G(\{n_j\},z_k)
=\beta_{k,0}^{-1}\prod_{\ell\ne k}(z-z_\ell).
\label{dhh}\end{equation}

We now know enough to calculate the matrix elements we need. Using (\ref{oep3}) we find
\begin{equation}
\fl\langle\Omega|{\bf E}_m^{-}\mathcal{T}_0(x_q,y_q)|\Omega\rangle=
-\beta_{m,0}
\sum_{ {\{0\le n_m\le N-1\}}\atop{n_1+\cdots+n_L=N}}
\omega^{\sum_{j} j n_j}\bar{G}(\{n_j\},z_{m})
\langle \{n_j\}|\mathcal{T}_0(x_q,y_q)|\Omega\rangle.
\label{oepto2}\end{equation}
Comparing (\ref{nto}) with (\ref{ck2}) for $Q=0$
and using $\sum_j N_j=\sum_j (L-j)n_j$, we find
\begin{eqnarray}
\langle \{n_i\}|\mathcal{T}_0(x_q,y_q)|\Omega\rangle
=N^{-{\frac 12}L}\omega^{-\sum_{j}jn_j}\sum_{a=0}^{N-1}
\prod_{j=1}^L\frac{y_p^N-x_q^N}
{y_p-x_q\omega^{a+N_j}}\nonumber\\
\quad=N^{1-{\frac 12}L}\omega^{-\sum_{j}j n_j}y_p^{rN}(1-x_q^N/y_p^N)
G(\{n_i\},(x_q/y_p)^N).
\label{nto2}\end{eqnarray}
Using (\ref{dh}), (\ref{hf}) and (\ref{nto2}), (\ref{oepto2}) becomes
\begin{eqnarray}
\fl\langle \Omega|{\bf E}_m^{-}\mathcal{T}_0(x_q,y_q)|\Omega\rangle
&&=-\beta_{m,0}y_p^{rN}(1-x_q^N/y_p^N)N^{1-\frac 12L}\,
{\mbox{\myeu h}}_{m}(x^N_q/y^N_p)
\nonumber\\
&&=-y_p^{rN}(1-x_q^N/y_p^N)N^{1-\frac 12L}\prod_{\ell\ne m}
({x_q^N}/{y_p^N}-z\vq_{\ell}).
\label{oepto3}\end{eqnarray}
{}From (\ref{transfer1a}), we find 
\begin{equation}
\fl\langle \Omega|\mathcal{T}_0(x_q,y_q)|\Omega\rangle=
\prod_{j=1}^r(X_j+Y_j),
\quad
\langle \Omega|{\bf E}_m^{-}\mathcal{T}_0(x_q,y_q)|\Omega\rangle
=Z_m\prod_{j\ne m}(X_j+Y_j).
\label{oeto}\end{equation}
Now equating the ratios given by (\ref{oeto}), (\ref{oto2})
with (I.44), and (\ref{oepto3}) we obtain
\begin{equation}
\frac{X_m+Y_m}{Z_m}=
\frac{x_q^N-y_p^N z\vq_m}{x_q^N-y_p^N}.
\label{ratio1}\end{equation}
Similarly, we use first (\ref{boem}) and (\ref{ntbo}), 
followed by (\ref{ck2}), (\ref{dh}) and (\ref{hf}), to find
\begin{eqnarray}
\fl\langle{\bar\Omega}|{\bf E}_m^{+}
\mathcal{T}_0(x_q,y_q)|{\bar\Omega}\rangle&&=
-\beta_{m,0}z_m(\mu_p x_p/\mu_q)^{rN}N^{-\frac 12L}\nonumber\\
\fl&&\qquad\times\sum_{ {\{0\le n_j\le N-1\}}\atop{n_1+\cdots+n_L=N}}
\bar{G}(\{n_j\},z_{m})\sum_{a=0}^{N-1}
\prod_{j=1}^L\frac{1-(y_q/x_p)^N}
{1-\omega^{-a-j-N'_j}y_q/x_p}\nonumber\\
\fl&&=-z_m(\mu_p x_p/\mu_q)^{rN}(1-y_q^N/x_p^N)
N^{1-\frac 12L}\prod_{\ell\ne m}
({y_q^N}/{x_p^N}-z\vq_{\ell}),
\label{boemtbo3}\end{eqnarray}
where $N'_j\equiv\sum_{\ell=1}^{j-1}(N-1-n_{\ell})=N(j-1)+1-j-N_j.$
Using
\begin{equation}
\fl\langle{\bar \Omega}|=(-1)^r\langle{ \Omega}|({\bf x}_0^+)^{(r)}
=\langle{ \Omega}|\prod_{j=1}^r {\bf E}^-_j,\quad
|{\bar \Omega}\rangle=({\bf x}_0^-)^{(r)}|{ \Omega}\rangle
=\prod_{j=1}^r {\bf E}^+_j|{ \Omega}\rangle,
\label{boo}\end{equation}
cf.\ (I.35) and (\ref{psi}), we obtain from (\ref{transfer1a}) 
\begin{equation}
\fl\langle{\bar \Omega}|\mathcal{T}_0(x_q,y_q)|{\bar\Omega}\rangle=
\prod_{j=1}^r(X_j-Y_j),
\quad
\langle{\bar\Omega}|{\bf E}_m^{+}
\mathcal{T}_0(x_q,y_q)|{\bar\Omega}\rangle
=Z_m\prod_{j\ne m}(X_j-Y_j).
\label{boetbo}\end{equation}
Hence, combining this with (\ref{botbo2}), (I.44) and
(\ref{boemtbo3}), one finds
\begin{equation}
\frac{X_m-Y_m}{Z_m}=
\frac{x_p^N-y_q^N z_m^{-1}}{x_p^N-y_q^N}.
\label{ratio2}\end{equation}

This shows that the rotations ${\bf \cal R}$ and ${\bf \cal S}$
in (\ref{transfer1}) can be written as direct products of $r$
two by two matrices ${\bf \cal R}_j$ and ${\bf \cal S}_j$, 
\begin{eqnarray}
\fl{\bf \cal S}=\prod_{j=1}^r {\bf \cal S}\vq_j,\quad
{\bf\cal R}=\prod_{j=1}^r {\bf\cal R}\vq_j,\quad
{\bf\cal S}\vq_j(A\vq_j-{\bf H}\vq_jB\vq_j)
{\bf\cal R}_j^{-1}=
X\vq_j-{\bf H}\vq_jY\vq_j+({\bf E}^+_j+{\bf E}^-_j)Z\vq_j.
\label{transfer1b} \end{eqnarray}
Since ${\bf \cal R}$ and ${\bf \cal S}$ must be independent
of $\lambda_q$, the left hand side of the last equation above is
linear in $\lambda^{-1}_q$, as seen from (\ref{abj}), and
so must its right hand side be. Thus, the ratios given in
(\ref{ratio1}), (\ref{ratio2}) and (I.2) yield the following
\begin{eqnarray}
\fl X_j+Y_j={\epsilon}\vq_jk(y_p^N z\vq_j-x_q^N)=
{\epsilon}_j[(1-k'\lambda_p)z_j-(1-k'\lambda^{-1}_q)],\nonumber\\
\fl Z_j={\epsilon}_j k(y_p^N -x_q^N)
 ={\epsilon}_j\lambda\vq_p\lambda^{-1}_q k(y_q^N-x_p^N )=
{\epsilon}_j k'(\lambda^{-1}_q-\lambda\vq_p),\nonumber\\
\fl X_j-Y_j={\epsilon}_j\lambda\vq_p\lambda^{-1}_q
k(y_q^N z_j^{-1}-x_p^N)
={\epsilon}_j[-k'\lambda\vq_p z_j^{-1}+
\lambda_q^{-1}(k'+\lambda_p z_j^{-1}-\lambda\vq_p)],
\label{XYZ1} \end{eqnarray}
We have inserted factors $\lambda\vq_p\lambda^{-1}_q$ in both the
numerator and the denominator of the second ratio in
(\ref{ratio2}) so that they become linear in $\lambda^{-1}_q$. Since
$\det({\bf \cal R}_j)=1$ and $\det({\bf \cal S}_j)=1$, the determinant
is invariant under the transformation in (\ref{transfer1b}), namely
\begin{equation}
X^2_j-Y^2_j-Z^2_j=A_j^2-B_j^2.
\label{invdet}\end{equation} 
Using (\ref{abj}), (\ref{theta}) and (\ref{tld}) in this order, we find
\begin{equation}
A_j^2-B_j^2=\rho^2 k^2(t_p^Nz_j-t_q^N)\big/(k'\lambda\vq_q),
\label{a2b2}\end{equation} 
whereas from equations on the left of (\ref{XYZ1}) and (I.5), we obtain
\begin{equation}
X^2_j-Y^2_j-Z^2_j={\epsilon}_j^2
k^2\lambda^{-1}_q\lambda\vq_p(z^{-1}_j-1)(t_p^Nz\vq_j-t_q^N).
\label{XYZ2}\end{equation}
Consequently we find the multiplicative constants
\begin{equation}
{\epsilon}_j^2={\bar\epsilon}_j^2\rho^2=
\rho^2/[k'(z_j^{-1}-1)\lambda\vq_p],\qquad
\rho^2=N^{1/r}k'/k^2.
\label{epsilon}\end{equation} 
Since the right hand side of (\ref{transfer1b}) is explicitly given
in (\ref{XYZ1}) and the $\lambda_q$ dependence in the left hand side
comes only from the $A_j$ and $B_j$ in (\ref{abj}), we conclude that
both sides are linear in $\lambda_q^{-1}$. Therefore, we can get two
equations by equating the coefficients of the constant and linear
terms of (\ref{transfer1b}) separately, i.e.
\begin{equation}
\fl{\bf \cal S}_j(\cosh\theta_j{\bf 1}-
\sinh\theta_j{\bf H}_j){\bf \cal R}^{-1}_j={\bf M},\quad
{\bf \cal S}_j(\cosh\theta_j{\bf 1}+
\sinh\theta_j{\bf H}_j){\bf \cal R}^{-1}_j=-{\bf N},
\label{soSRj}\end{equation}
where the matrices ${\bf M}$ and ${\bf N}$ depend on $j$ and
have elements
\begin{eqnarray}
\fl m_{11}=-{\bar\epsilon}_j k'\lambda_p/z_j,\qquad
m_{12}=m_{21}=-{\bar\epsilon}_j k'\lambda_p,\qquad
m_{22}={\bar\epsilon}_j (z_j-1-k'z_j\lambda_p),\nonumber\\
\fl n_{11}={\bar\epsilon}\vq_j (z_j^{-1}\lambda\vq_p-\lambda\vq_p+k'),
\qquad n_{12}=n_{21}=n_{22}={\bar\epsilon}_j k',
\qquad k'(z_j^{-1}-1)\lambda\vq_p{\bar\epsilon}_j^{\,2}=1.
\label{mnij} \end{eqnarray} 
It is straightforward to solve for ${\bf \cal R}_j$ and
${\bf \cal S}_j$. We shall leave the details to the appendix
and present the results here. We find that 
\begin{eqnarray}
{\bf \cal S}_j=\halft(s_{11}+s_{22}){\bf 1}+
\halft(s_{11}-s_{22}){\bf H}\vq_j+
s_{12}{\bf E}^+_j+s_{21}{\bf E}^-_j,
\label{Sj} \end{eqnarray}
with
\begin{eqnarray}
s_{22}=\biggl(\frac{m_{22}\rme^{\theta_j}+
n_{22}\rme^{-\theta_j}}{2\sinh 2\theta_j}\biggr)^{1/2},\quad
&&
s_{12}=\frac{m_{12}\rme^{\theta_j}+
n_{12}\rme^{-\theta_j}}{m_{22}\rme^{\theta_j}+
n_{22}\rme^{-\theta_j}}\,s_{22},\nonumber\\
s_{21}=\frac{\rme^{-2\theta_j}-k'}{2s_{12}\sinh 2\theta_j},\quad
&& s_{11}=\frac{\rme^{2\theta_j}-k'}{2s_{22}\sinh 2\theta_j},
\label{Sij} \end{eqnarray}
where the $\theta_j$ are defined in (\ref{theta}),
while ${\bf \cal R}_j$ is the transpose of the inverse
of ${\bf \cal S}_j$, that is
\begin{eqnarray}
r_{22}=s_{11},\quad r_{21}=-s_{12},\quad r_{12}=-s_{21},
\quad r_{11}=s_{22}.
\label{Rij} \end{eqnarray}

We now show that the ${\bf \cal R}$ and ${\bf \cal S}$ in equation
(\ref{transfer23}) for the other two transfer matrices are identical.
To calculate the matrix elements of ${\hat\mathcal{T}}_Q(y_q,x_q)$ we
use (\ref{ctcbt}) and (I.4), while using an equation like (\ref{wwb}),
but with the replacements $p\leftrightarrow p'$, $q\leftrightarrow q'$,
(or $x_r\leftrightarrow y_r$ and $\mu_r\rightarrow\mu_r^{-1}$ with
$r=p$, $q$).
Similar to (\ref{otrans1}) and (\ref{nntrans1}), we may then write 
\begin{eqnarray}
&\langle\{n'_i\}|{\hat\mathcal{T}}_Q(y_q,x_q)|\{n_i\}\rangle=
N^{-{\frac 12}L}
\sum_{a=0}^{N-1}\omega^{-Qa}\nonumber\\
&\qquad\times\prod_{j=1}^L\Biggl[{\frac{(x_p^N-y_q^N)
\omega^{a-N\vq_{j}+N'_j}}
{x_p-y_q\omega^{a-N\vq_{j}+N'_j}}}
{\Bigl(\frac {\mu_q}{\mu_p}\Bigr)}^{n'_j}\prod^{n'_j}_{\ell=1}
\frac{x_q-y_p\omega^{\ell+a-N\vq_{j}+N'_j}}
{x_p-y_q\omega^{\ell+a-N\vq_{j}+N'_j}}\Biggr].
\label{ntrans3}\end{eqnarray}
Here, when $n_i=0$ or $n_i=N-1$, the above expressions are related to
the functions defined in (\ref{ck2}) and (\ref{chk2}), i.e.
\begin{eqnarray}
\fl\langle\Omega|\hat{\mathcal{T}}_0(y_q,x_q)|\{n_i\}\rangle=
N^{1-{\frac 12}L}\omega^{\sum_{j=1}^Lj n_j}x_p^{rN}(1-y_q^N/x_p^N)
\bar G(\{n_i\},(y_q/x_p)^N),\label{nto3}\\
\fl\langle\bar\Omega|\hat{\mathcal{T}}_0(y_q,x_q)|
\{N\!-\!1\!-\!n_i\}\rangle=
N^{1-{\frac 12}L}(y_p\mu_q/\mu_p)^{rN}(1-x_q^N/y_p^N)
\bar G(\{n_i\},(x_q/y_p)^N).
\label{nto4}\end{eqnarray}
Combining this with (\ref{emo2}), (\ref{epbo}) and (\ref{dhh}),
and also noting that $(1-z)\bar G(\{0\},z)=P(z)$ for $n_j\equiv0$,
we can derive
\begin{eqnarray}
\fl\frac{\langle \Omega|{\hat\mathcal{T}}_0(y_q,x_q)|\Omega\rangle}
{\langle \Omega|{\hat\mathcal{T}}_0(y_q,x_q){\bf E}_m^{+}|
\Omega\rangle}
=-\frac{x_p^N-y_q^N z_m^{-1}}{x_p^N-y_q^N},\quad
\frac{\langle{\bar\Omega}|{\hat\mathcal{T}}_0(y_q,x_q)|
{\bar\Omega}\rangle}
{\langle{\bar \Omega}|{\hat\mathcal{T}}_0(y_q,x_q){\bf E}^{-}_m|
{\bar\Omega}\rangle}
=-\frac{x_q^N-y_p^N z\vq_m}{x_q^N-y_p^N}.
\label{ratio3}\end{eqnarray}
Setting $x_q\leftrightarrow y_q$ in (\ref{ratio3}), we find 
\begin{eqnarray}
\fl\frac{\langle \Omega|{\hat\mathcal{T}}_0(x_q,y_q)|\Omega\rangle}
{\langle \Omega|{\hat\mathcal{T}}_0(x_q,y_q){\bf E}_m^{+}|
\Omega\rangle}=
-\frac{x_p^N-x_q^N z_m^{-1}}{x_p^N-x_q^N},\quad
\frac{\langle{\bar\Omega}|{\hat\mathcal{T}}_0(x_q,y_q)|
{\bar\Omega}\rangle}
{\langle{\bar \Omega}|{\hat\mathcal{T}}_0(x_q,y_q){\bf E}^{-}_m|
{\bar\Omega}\rangle}
=-\frac{y_q^N-y_p^N z\vq_m}{y_q^N-y_p^N}.
\label{ratio4}\end{eqnarray}
Similar to (\ref{transfer1a}) and (\ref{transfer1b}), we
find that (\ref {transfer23}) can be written as
\begin{eqnarray}
\fl{\hat{\cal T}}_0(y_q,x_q)=
\prod_{j=1}^r[{\bar X}\vq_j-{\bf H}\vq_j {\bar Y}\vq_j+
({\bf E}^+_j +{\bf E}^-_j){\bar Z}\vq_j]=
\prod_{j=1}^r{\bf \cal R}\vq_j
({\bar A}\vq_j-{\bf H}\vq_j{\bar B}\vq_j){\bf \cal S}^{-1}_j,
\label{transfer2b}\\
\fl{\hat{\cal T}}_0(x_q,y_q)=
\prod_{j=1}^r[X'_j-{\bf H}\vq_j Y'_j+
({\bf E}^+_j +{\bf E}^-_j)Z'_j]=
\prod_{j=1}^r{\bf \cal R}\vq_j
( A\vq_j-{\bf H}\vq_j B\vq_j){\bf \cal S}^{-1}_j.
\label{transfer3b}\end{eqnarray}

Since ${\bar A_j}$ and ${\bar B}_j$ are obtained from the $A_j$
and $B_j$ in (\ref{abj}) by letting $\lambda_q^{-1}\to\lambda\vq_q$
as follows from (I.2) for $\lambda\vq_q\equiv\mu_q^N$, they are
linear in $\lambda_q$, so that, similar to (\ref{XYZ1}),
(\ref{ratio3}) yields
\begin{eqnarray}
\fl{\bar X}_j+{\bar Y}_j=
{\epsilon}_j \lambda_p k(x_p^N -y_q^N z^{-1}_j),\quad
{\bar Z}_j=-{\epsilon}_j\lambda_p k(x_p^N -y_q^N)
 =-{\epsilon}_j\lambda_q k(x_q^N-y_p^N ),\nonumber\\
{\bar X}_j-{\bar Y}_j={\epsilon}_j\lambda_q k(x_q^N-y_p^N z\vq_j),
\label{bXYZ} \end{eqnarray}
while each term in the product of (\ref{transfer3b}) is linear
in $\lambda_q^{-1}$, so that (\ref{ratio4}) leads to
\begin{eqnarray}
\fl X'_j+Y'_j={\epsilon}_j \lambda_p k(x_p^N -x_q^N z^{-1}_j),\quad
Z'_j=-{\epsilon}_j\lambda_p k(x_p^N -x_q^N)
 =-{\epsilon}_j\lambda^{-1}_q k(y_q^N-y_p^N ),\nonumber\\
X'_j-Y'_j={\epsilon}_j\lambda^{-1}_q k(y_q^N-y_p^N z\vq_j).
\label{pXYZ} \end{eqnarray}
By equating separately the coefficients of the terms constant and
linear in $\lambda_q$ of the $j$th factors of the products on both
sides of (\ref{transfer2b}), we find
\begin{equation}
\fl{\bf \cal R}_j(\cosh\theta_j{\bf 1}-
\sinh\theta_j{\bf H}_j){\bf \cal S}^{-1}_j=-{\bf N}^{-1},\quad
{\bf \cal R}_j(\cosh\theta_j{\bf 1}+
\sinh\theta_j{\bf H}_j){\bf \cal S}^{-1}_j={\bf M}^{-1},
\label{soSRf}\end{equation}
where the elements of the matrices ${\bf M}$ and ${\bf N}$ were
given in (\ref{mnij}) and for their inverses we used
$\det{\bf M}=\det{\bf N}=1$. From (\ref{transfer3b}) we get the
same equation (\ref{soSRf}), as this case differs only by the
interchange $q\leftrightarrow q'$,
$\lambda_q\leftrightarrow1/\lambda_q$. Taking the inverses of the
two equations in (\ref{soSRf}) we recover (\ref{soSRj}). This proves
that we can indeed use the identical ${\bf \cal R}_j$
and ${\bf \cal S}_j$.

To summarize our results, in (\ref{transfer1a}) and (\ref{transfer3b})
we have expressed the transfer matrices of the superintegrable chiral
Potts model for the $Q=0$ ground state sector in terms of the generators
of ${\mathfrak{sl}}_2$ algebras, with constants $X_j, Y_j, Z_j$ and
$X'_j, Y'_j, Z'_j$ given explicitly in (\ref{XYZ1}) and (\ref{pXYZ}),
where $j=1,\ldots,r$ with $r=(N-1)L/N$ and $L$ a multiple of $N$. The
corresponding two sets of $2^r$ eigenstates for the transfer matrices
are given in (\ref{evector}) and (\ref{eigeneq}) with the rotations
${\bf \cal R}$ and ${\bf \cal S}$ explicitly given in (\ref{transfer1b})
and (\ref{Sj}) to (\ref{Rij}).

\section*{Acknowledgements}
The authors thank Professor Rodney Baxter for his valuable comments and
criticism. The suggestions and interest of Professor Tetsuo Deguchi are
also deeply appreciated.
We thank our colleagues and the staff at the Centre for Mathematics and
its Applications (CMA) and at the Department of Theoretical Physics
(RSPE) of Australian National University for their generous support and
hospitality.

\appendix
\section{\boldmath${\bf H}_m|\Omega\rangle=-|\Omega\rangle$ and
\boldmath$({\bf E}^+_m)^2|\Omega\rangle=0$}
After applying $x_0^+$ to (I.42), we find, using (I.37) and
$\mathbf{h}\vq_n=x_0^+x_n^--x_n^-x_0^+$ in (I.34),
\begin{equation}
n\Lambda_{n}=\sum_{j=1}^n
d_j \Lambda_{n-j},\quad \mbox{where}\quad
{\bf h}_n|\Omega\rangle=d_n|\Omega\rangle.
\label{dlam}\end{equation}
Similar to what is done in \cite{Degu1}, we multiply both side by $z^n$
and then sum over $n$ from 1 to $\infty$. Next, noting $\Lambda_n=0$
for $n>r$, we divide the result by $P(z)$ and obtain
\begin{equation}
z{\frac d{dz}}\ln P(z)=\sum_{n=1}^{\infty} d_n z^n=
\sum_{j=1}^r {\frac z{z-z_j}}
=-\sum_{n=1}^{\infty} z^n \sum_{j=1}^r z_j^{-n}.
\label{dp}\end{equation}
This shows $d_n=- \sum_{j=1}^r z_j^{-n}$. Using (\ref{vdm}) we obtain
\begin{equation}
\sum_{n=0}^{r-1}\beta_{m^\ast,n}d_{n+k}=
-\sum_{j=1}^r\sum_{n=0}^{r-1}\beta\vq_{m^\ast,n} z_{j^\ast}^{n+k}=
-\sum_{j=1}^r\delta\vq_{m,j}z_j^{-k}
=-z_m^{-k}.
\label{bd}\end{equation}
{}From (\ref{eeh}), we find
\begin{equation}
{\bf H}_m|\Omega\rangle=
\sum_{n=0}^{r-1}\beta_{m^\ast,n}d_n|\Omega\rangle
=-|\Omega\rangle.
\label{hop}\end{equation}
From (\ref{comEH}) we see that ${\bf E}_m^{\pm}$ and ${\bf H}_m$
generate a finite-dimensional representation of $\frak{sl}(2)$.
We also have ${\bf E}_m^-|\Omega\rangle=0$. Thus this representation
is spin-$\frac12$, proving $({\bf E}_m^{+})^2|\Omega\rangle=0$.
\section{Results for \boldmath${\bf E}^+_j{\bf E}^+_m|\Omega\rangle$}

As the identities in (\ref{eeh}) hold for the entire eigenspace,
whereas the identity in (\ref{emo}) holds only for the ground state, we
use (\ref{eeh}) with $\ell=0$ for ${\bf E}_j^{+}$ and (\ref{emo})
for ${\bf E}_m^{+}$ to find
\begin{equation}
{\bf E}_j^{+}{\bf E}_m^{+}|\Omega\rangle=
\beta_{m,0}\sum_{\ell=1}^r z_m^{\ell}
\sum_{\ell'=0}^{r-1} \beta\vq_{j^\ast,\ell'}{\bf x}_{\ell'}^-
({\bf x}_0^+)^{(\ell-1)}
({\bf x}_1^-)^{(\ell)}|\Omega\rangle.
\label{eeo}\end{equation}
As $({\bf x}_m^{\pm})^{(n)}\equiv({\bf x}_m^{\pm})^n/n!$ for $n\ge0$
and $({\bf x}_m^{\pm})^{(n)}\equiv0$ for $n<0$,
it is easy to prove by induction that the following relations hold for
the entire eigenspace,
\begin{eqnarray}
&[({\bf x}_0^+)^{(j)},{\bf x}_k^-]=
({\bf x}_0^+)^{(j-1)}{\bf h}\vq_k-
{\bf x}_k^+({\bf x}_0^+)^{(j-2)}, \nonumber\\
&[{\bf x}_k^+,({\bf x}_1^-)^{(j)}]=
({\bf x}_1^-)^{(j-1)}{\bf h}\vq_{k+1}+
{\bf x}_{k+2}^-({\bf x}_1^-)^{(j-2)},\nonumber\\
&[{\bf h}\vq_k,({\bf x}_0^+)^{(j)}]=
-2{\bf x}_k^+({\bf x}_0^+)^{(j-1)},
\quad [{\bf h}\vq_k,({\bf x}_1^-)^{(j)}]=
2{\bf x}_{k+1}^-({\bf x}_0^-)^{(j-1)},
\label{commuta}\end{eqnarray}
which are very similar to those in~\cite{Degu1}.
Using (\ref{commuta}) and (\ref{dlam}) we interchange the order
of operations in (\ref{eeo}), i.e.
\begin{eqnarray}
\fl{\bf x}_{\ell'}^-({\bf x}_0^+)^{(\ell-1)}
({\bf x}_1^-)^{(\ell)}|\Omega\rangle=
\{({\bf x}_0^+)^{(\ell-1)}({\bf x}_1^-)^{(\ell)}{\bf x}_{\ell'}^-
-({\bf x}_0^+)^{(\ell-2)}[2({\bf x}_1^-)^{(\ell-1)}
{\bf x}_{\ell'+1}^-\nonumber\\
+({\bf x}_1^-)^{(\ell)}d_{\ell'}]
+({\bf x}_0^+)^{(\ell-3)}[({\bf x}_1^-)^{(\ell-2)}
{\bf x}_{\ell'+2}^-+
({\bf x}_1^-)^{(\ell-1)}d_{\ell'+1}]\}
|\Omega\rangle.
\label{xmpm}\end{eqnarray}
Substituting (\ref{xmpm}) into (\ref{eeo}), and using (\ref{eeh}) and
(\ref{bd}), we get
\begin{eqnarray}
&\fl{\bf E}_j^{+}{\bf E}_m^{+}|\Omega\rangle&=
\beta_{m,0}\Bigl\{\sum_{\ell=1}^r z_m^{\ell}
({\bf x}_0^+)^{(\ell-1)}({\bf x}_1^-)^{(\ell)}{\bf E}_j^{+}
|\Omega\rangle\nonumber\\
&\fl&\qquad\qquad-\sum_{\ell=2}^r z_m^{\ell}({\bf x}_0^+)^{(\ell-2)}
\bigl[({\bf x}_1^-)^{(\ell-1)}2z_j^{-1}
{\bf E}_j^{+}-({\bf x}_1^-)^{(\ell)}\bigr]|\Omega\rangle\nonumber\\
&\fl&\qquad\qquad+\sum_{\ell=3}^r z_m^{\ell}({\bf x}_0^+)^{(\ell-3)}
\bigl[({\bf x}_1^-)^{(\ell-2)}z^{-2}_j
{\bf E}_j^{+}-({\bf x}_1^-)^{(\ell-1)}z^{-1}_j\bigr]|\Omega\rangle
\Bigr\}\nonumber\\
&\fl&=\beta_{m,0}\Bigl\{
(1-2z_m/z_j+z_m^2/z_j^2)\sum_{\ell=1}^{r-1} z_m^{\ell}
({\bf x}_0^+)^{(\ell-1)} ({\bf x}_1^-)^{(\ell)}{\bf E}_j^{+}
|\Omega\rangle\nonumber\\
&\fl&\qquad\qquad+(1-z_m/z_j)
\sum_{\ell=2}^{r} z_m^{\ell}({\bf x}_0^+)^{(\ell-2)}
({\bf x}_1^-)^{(\ell)}|\Omega\rangle\nonumber\\
&\fl&\qquad\qquad+ z_m^{r}({\bf x}_0^+)^{(r-1)}
({\bf x}_1^-)^{(r)}{\bf E}_j^{+}|\Omega\rangle\nonumber\\
&\fl&\qquad\qquad-z_m^{r+1}({\bf x}_0^+)^{(r-2)}
[ z^{-2}_j({\bf x}_1^-)^{(r-1)}{\bf E}_j^{+}
-({\bf x}_1^-)^{(r)}z^{-1}_j]|\Omega\rangle\Bigr\}.
\label{eeo1}\end{eqnarray}
To arrive at the second equality, we changed the summations in
the first equality replacing $\ell-1\to\ell$ both in the first part
of the second sum and in the second part of the third sum, and
$\ell-2\to\ell$ in the first part of the third sum.
If we can use the identities,
\begin{equation}
({\bf x}_1^-)^{(r)}{\bf E}_j^{+}|\Omega\rangle=0,\quad
({\bf x}_1^-)^{(r-1)}{\bf E}_j^{+}
|\Omega\rangle=z_j({\bf x}_1^-)^{(r)}|\Omega\rangle,
\label{iden}\end{equation}
we may simplify (\ref{eeo1}) to
\begin{eqnarray}
{\bf E}_j^{+}{\bf E}_m^{+}|\Omega\rangle&=&
\beta_{m,0}\Bigl\{
(1-z_m/z_j)^2\sum_{\ell=1}^{r-1}
z_m^{\ell}({\bf x}_0^+)^{(\ell-1)}
({\bf x}_1^-)^{(\ell)}{\bf E}_j^{+}|\Omega\rangle\nonumber\\
&&+(1-z_m/z_j)\sum_{\ell=2}^{r} z_m^{\ell}({\bf x}_0^+)^{(\ell-2)}
({\bf x}_1^-)^{(\ell)}|\Omega\rangle
\Bigr\}.
\label{eeo2}\end{eqnarray}
This is clearly consistent with $({\bf E}_m^{+})^2|\Omega\rangle=0$.

{}From (\ref{emo2}), (I.35) and (\ref{oefo}), it is easy to see that
the first identity in (\ref{iden}) holds. To prove the second identity,
we use (\ref{xmo1}) with $n_j\equiv N-1$ to find
\begin{equation}
({\bf x}_1^-)^{(r)}|\Omega\rangle=
\omega^{\frac 12 L(L+1)}|{\bar\Omega}\rangle.
\label{x1ro}\end{equation}
Next use (\ref{emo2}), (I.35) with $\nu_m=N-1-n_m$, (\ref{oefo}),
and finally (\ref{sfactor}) with $n=0$, to find
\begin{equation}
\fl({\bf x}_1^-)^{(r-1)}{\bf E}_j^{+}|\Omega\rangle=
\omega^{\frac 12 L(L+1)}|{\bar\Omega}\rangle
(-z_j\beta_{j,0})\sum_{ {\{0\le n_m\le N-1\}}\atop{n_1+\cdots+n_L=N} }
\omega^{-\sum_m m n_m}G(\{n_m\},z_j).
\label{xr-1e}\end{equation}
Multiplying both sides of (\ref{dhh}) by $z^{-r+1}$ and then taking
the limit $z\equiv t^N\to\infty$, using (\ref{chk2}) to obtain the
limit $t^{-1}\to 0$ of $t^{-Nr+N}{\bar G}$, where
$\sum_m\bar N_m=\sum_m (m-1)n_m$, we find 
\begin{equation}
\lim_{z\to\infty}z^{-r+1}{\bar{\mbox{\myeu h}}}_k(z)=
-\sum_{ {\{0\le n_j\le N-1\}}\atop{n_1+\cdots+n_L=N}}
\omega^{-\sum_m m n_m}G(\{n_j\},z_k)
=-\beta_{k,0}^{-1}.
\label{dhhlim}\end{equation}
Substituting this equation into (\ref{xr-1e}), we show that the
second identity in (\ref{iden}) holds.

\section{The rotations \boldmath${\bf R}_j$ and ${\bf S}_j$}
We rewrite (\ref{soSRj}) as
\begin{equation}
\fl{\bf \cal S}_j(\cosh\theta_j{\bf 1}-\sinh\theta_j{\bf H}_j)=
{\bf M}{\bf \cal R}_j,\quad
{\bf \cal S}_j(\cosh\theta_j{\bf 1}+\sinh\theta_j{\bf H}_j)=
-{\bf N}{\bf \cal R}_j,
\label{soSRj1}\end{equation}
where $\det({\bf M})=\det({\bf N})=1$. We multiply the first equation
with ${\bf M}^{-1}$ and the second equation with $-{\bf N}^{-1}$,
so that the right-hand sides of both equations become ${\bf \cal R}_j$
and the left-hand sides must be equal. This way we obtain a matrix
equation for ${\bf \cal S}_j$. More explicitly its elements
$s_{ij}$ can be shown to satisfy the following equations,
\begin{eqnarray}
&&s_{21}(m_{11}\rme^{-\theta_j}+n_{11}\rme^{\theta_j})=
s_{11}(m_{21}\rme^{-\theta_j}+n_{21}\rme^{\theta_j}),\nonumber\\
&&s_{21}(m_{12}\rme^{-\theta_j}+n_{12}\rme^{\theta_j})=
s_{11}(m_{22}\rme^{-\theta_j}+n_{22}\rme^{\theta_j}),\nonumber\\
&&s_{22}(m_{11}\rme^{\theta_j}+n_{11}\rme^{-\theta_j})=
s_{12}(m_{21}\rme^{\theta_j}+n_{21}\rme^{-\theta_j}),\nonumber\\
&&s_{22}(m_{12}\rme^{\theta_j}+n_{12}\rme^{-\theta_j})=
s_{12}(m_{22}\rme^{\theta_j}+n_{22}\rme^{-\theta_j}).
\label{Sj2}\end{eqnarray}
These equations are consistent if and only if
\begin{eqnarray}
(m_{11}\rme^{\mp\theta_j}+n_{11}\rme^{\pm\theta_j})
(m_{22}\rme^{\mp\theta_j}+n_{22}\rme^{\pm\theta_j})\nonumber\\
\qquad=
(m_{21}\rme^{\mp\theta_j}+n_{21}\rme^{\pm\theta_j})
(m_{12}\rme^{\mp\theta_j}+n_{12}\rme^{\pm\theta_j}).
\label{consist}\end{eqnarray}
We can simplify (\ref{consist}) using $m_{12}=m_{21}$ and
$n_{12}=n_{21}$ found in (\ref{mnij}) and  again using
$\det({\bf M})=\det({\bf N})=1$. We find that the two conditions
become
\begin{eqnarray}
2\cosh2\theta_j=2m_{12}n_{12}-m_{11}n_{22}-m_{22}n_{11}.
\label{consist2}\end{eqnarray}
Next, using (\ref{tld}) and (\ref{theta}), we can show
\begin{equation}
\lambda\vq_p+\lambda_p^{-1}=
2z_j^{-1}\cosh 2\theta_j-(k'+{k'}\vphantom{z}^{-1})(z_j^{-1}-1).
\label{lamthe}\end{equation}
With the help of (\ref{mnij}) and (\ref{lamthe}) we can then prove
that (\ref{consist2}) holds.

Moreover, from $\det{\bf \cal S}_j=1$ we calculate $s_{11}s_{22}$
eliminating first $s_{12}s_{12}$ using (\ref{Sj2}) and (\ref{mnij}),
and then substituting $z_j$ from (\ref{lamthe}). This way we find
\begin{equation}
s_{11}s_{22}=\frac {e^{2\theta_j}-k'}{2\sinh 2\theta_j},\qquad
s_{12}s_{21}=\frac {e^{-2\theta_j}-k'}{2\sinh 2\theta_j}.
\label{Sj3}\end{equation}
Following nearly identical steps, we obtain from (\ref{soSRj1})
consistent equations similar to (\ref{Sj2}) and (\ref{Sj3}) for
the matrix elements $r_{ij}$ of ${\bf \cal R}_j$. From these results
we conclude
\begin{eqnarray}
\frac {r_{22}}{r_{12}}=-\frac {s_{11}}{s_{21}},\quad
\frac {r_{11}}{r_{21}}=-\frac {s_{22}}{s_{12}},\qquad
{{ r_{22}r_{11}=s_{11}s_{22},}\atop{ r_{12}r_{21}=s_{12}s_{21}.}}
\label{Rj2}\end{eqnarray}

Only six of the equations in (\ref{Sj2}), (\ref{Sj3}) and (\ref{Rj2})
are independent. At this point we may choose $s_{22}$ and $r_{22}$
as free parameters, but then the other six elements are uniquely
determined. However, there is one more relation.
From direct multiplication of the matrices in (\ref{soSRj}), we find 
\begin{eqnarray}
m_{11}=\rme^{-\theta_j}s_{11}r_{22}-\rme^{\theta_j}s_{12}r_{21},\quad
n_{11}=-\rme^{\theta_j}s_{11}r_{22}+\rme^{-\theta_j}s_{12}r_{21}.
\label{SRj2}\end{eqnarray}
Eliminating $s_{12}r_{21}$ from these, (and similarly starting with
$m_{22}$ and $n_{22}$), we find
\begin{eqnarray}
s_{11}r_{22}=-\frac{m_{11}\rme^{-\theta_j}+
n_{11}\rme^{\theta_j}}{2\sinh 2\theta_j},\qquad
s_{22}r_{11}=\frac{m_{22}\rme^{\theta_j}+
n_{22}\rme^{-\theta_j}}{2\sinh 2\theta_j}.
\label{RSj3}\end{eqnarray}
Let us define
\begin{eqnarray}
T\vq_{ik}\equiv m_{ik}\rme^{-\theta_j}+n_{ik}\rme^{\theta_j}=T\vq_{ki},
\quad
T^*_{ik}\equiv m_{ik}\rme^{\theta_j}+n_{ik}\rme^{-\theta_j}=T^*_{ki}.
\label{TTs}\end{eqnarray}
Then it can be easily verified, substituting (\ref{mnij})
and eliminating $z_j$ from (\ref{lamthe}), that
\begin{eqnarray}
T\vq_{12}T^*_{12}=-(\rme^{-2\theta_j}-k')(\rme^{2\theta_j}-k'),
\nonumber\\
T\vq_{11}T^*_{22}=-(\rme^{2\theta_j}-k')^2,\quad
T\vq_{22}T^*_{11}=-(\rme^{-2\theta_j}-k')^2.
\label{Tid}\end{eqnarray}
Using (\ref{Sj3}) and (\ref{Tid}), we may rewrite (\ref{RSj3}) as
\begin{equation}
\frac {r_{22}}{s_{22}}=-\frac {T_{11}}{e^{2\theta_j}-k'}=
\frac{e^{2\theta_j}-k'} {T^*_{22}}.
\label{SRj4}\end{equation}
This shows that there is only one free parameter $s_{22}$, and we
choose it so that $r_{11}=s_{22}$. This choice
yields the solution given in (\ref{Sij}) and (\ref{Rij}).


\end{document}